\documentclass[nomathfonts]{aipproc} 
\layoutstyle{6s}

\usepackage{makeidx}
\usepackage{eurosym}
\usepackage{amsfonts}
\usepackage[latin1]{inputenc}
\usepackage{times}
\usepackage[T1]{fontenc}

\usepackage[dvips]{color}
\usepackage{xspace}
\usepackage[dvips, colorlinks,linkcolor=webgreen,filecolor=webbrown,citecolor=webgreen,
breaklinks=true, urlcolor=webbrown,pdfstartview=FitH ]{hyperref}
\usepackage{amsmath}
\usepackage{amssymb}

\newtheorem{theorem}{Theorem}

\newtheorem{definition}[theorem]{Definition}

\newtheorem{lemma}[theorem]{Lemma}
\newtheorem{notation}[theorem]{Notation}

\newtheorem{remark}[theorem]{Remark}

\DeclareMathAlphabet{\mathbfit}{OT1}{ptm}{bx}{it}

\definecolor{webgreen}{rgb}{0,.5,0}
\definecolor{webbrown}{rgb}{.6,0,0}
\definecolor{webyellow}{rgb}{0.98,0.92,0.73}
\definecolor{webgray}{rgb}{.753,.753,.753}
\definecolor{webblue}{rgb}{0,0,.8}
\newlength{\coco}
 \divide \coco 2

\title{An amended MaxEnt formulation for deriving Tsallis factors, and associated issues
\footnotetext{This work was presented at MaxEnt2006, the 26$^\mathrm{th}$  International Workshop on Bayesian Inference and Maximum Entropy Methods in Science and Engineering CNRS, Paris, France, July 8-13, 2006.}
}
\author{\href{mailto:jf.bercher@esiee.fr}{Jean-François Bercher}}
 {
  address = {\'Equipe Signal et Information,\linebreak 
  Dept. Modélisation et Simulation  \linebreak  
  ESIEE, Noisy-le-Grand, France},
  email = {jf.bercher@esiee.fr}
 }

\date{}

\begin{abstract}

An amended MaxEnt formulation for systems displaced from the
conventional MaxEnt equilibrium is proposed. This formulation involves the minimization of
the Kullback-Leibler divergence to a reference $Q$ (or maximization of Shannon
$Q$-entropy), subject to a constraint that implicates a second reference
distribution $P_{1}$ and tunes the new equilibrium. In this setting, the
equilibrium distribution is the generalized escort distribution associated to
$P_{1}$ and $Q$. The account of an additional constraint, an observable given
by a statistical mean, leads to the maximization of Rényi/Tsallis $Q$-entropy
subject to that constraint. Two natural scenarii for this
observation constraint are considered, and the classical and generalized constraint of
nonextensive statistics are recovered. The solutions to the
maximization of Rényi $Q$-entropy subject to the two types of constraints are derived.
These optimum distributions, that are Levy-like distributions, are
self-referential. We then propose two \ `alternate' (but effectively
computable) dual functions, whose maximizations enable to identify the optimum parameters. Finally,
a duality between solutions and the underlying Legendre structure are presented.
\end{abstract}

\begin{document}
\maketitle

\section{Introduction}

The    formalism    of    nonextensive    statistical    mechanics
\cite{Tsallis1999, Tsallis2002} leads  to a generalized  Boltzmann
factor in  the form  of a  Tsallis distribution  (or factor)  that
depends on an entropic index and recovers the classical  Boltzmann
factor   as  a   special  limit   case  \cite{Tsallis1999}.   
This distribution  is of  high
interest in many physical systems since it enables to model  power-law 
phenomena.
% and was applied  successfully in a wide variety  of fields.
In a  wide variety  of
fields, experiments, numerical results and analytical  derivations
fairly agree with the description by a Tsallis distribution. 

Tsallis' distributions (sometimes called Levy distributions) are derived by maximization
of Tsallis entropy \cite{Tsallis1988}, 
under suitable constraints. The present formulation is as follows: maximize Tsallis' entropy
\begin{equation}
T_{\alpha}(P)=\frac{1}{1-\alpha}\left[  \int P(x)^{\alpha}dx-1\right]  ,
\end{equation}
subject to 
\begin{equation}
m=\int x P^*(x)dx \mbox{~~with~~}  P^*(x)=\frac{P(x)^{\alpha}}{\int P(x)^{\alpha} dx},
\label{escort1}
\end{equation}
where the mean constraint is called a `generalized' mean constraint in the nonextensive litterature, and $P^*(x)$ is called the `escort' distribution. This formulation was preferred to the simple maximization with a classical mean constraint $m=\int x P(x)dx$ because of mathematical difficulties. The solution is given in the litterature as  
\begin{equation}
P(x) = \frac{1}{Z} \left( 1- \frac{(1-\alpha)\beta}{Z^{1-\alpha}} (x-m) \right)^{\frac{1}{1-\alpha}},
\label{FistTsal}
\end{equation}
where $Z$ is a partition function.

Of course, these distributions  do not coincide with those derived
by conventionnal  MaxEnt and  consequently will  not be  justified
from a probabilistic point of  view, because of the uniqueness  of
the    rate   function    in   the    large   deviations    theory
\cite{LaCour2000b,  Grendar2004}.  Furthermore,  the  status   and
interest of generalized  expectations and of  escort distributions
is  unclear.  Last,  it   is  apparent  that  the   expression  of
distribution  (\ref{FistTsal})  is  implicit,  so  that  both  its
manipulation and  determination of  its parameter  $\beta$ will be
difficult. 

However, in view of the success of nonextensive statistics,  there
should exist a probabilistic setting that provides a justification
for the  maximization of  Tsallis entropy.  There are  now several
indications   that   results   of   nonextensive   statistics  are
physically relevant for partially equilibrated or  nonequilibrated
systems, with a stationary state characterized by fluctuations  of
an intensive parameter \cite{Beck2004a,  Wilk2000}; for  instance,
the Tsallis factor  is obtained from  the Boltzmann-Gibbs' if  the
inverse   of   temperature  fluctuates   according   to  a   gamma
distribution.\\

In this  paper, I  present a   framework for  the maximization  of
Rényi/Tsallis  $Q-$entropy,  that  leads  to  the  so-called  Levy
distribution (or  Tsallis factor).  
The Rényi information divergence, the opposite of Rényi $Q$-entropy,
is given by
 \begin{equation}
 D_{\alpha}(P||Q)=\frac{1}{\alpha-1}\log\int P(x)^{\alpha}Q(x)^{1-\alpha}dx,
 \label{Renyi}
 \end{equation}
where $\alpha$  is a real parameter called  the entropic index. 
Using L'Hospital's rule, the Kullback-Leibler divergence is recovered for $\alpha\rightarrow1$
 \begin{equation}
 D(P||Q)=\int P(x)\log\frac{P(x)}{Q(x)}dx.
 \end{equation}
Its opposite is the Shannon $Q-$entropy, the correct, coordinate invariant, extension of the
 classical Shannon entropy to the continuous case \cite{Jaynes1963}. This divergence can be interpreted as
 a \textquotedblleft distance\textquotedblright\ between two distributions. Rényi and Tsallis $Q$-entropies are related by a simple monotonic function.
 Therefore, their maximization under the same constraint lead to the same distribution.

% and the Shannon $Q$-entropy is recovered in the limit $\alpha\rightarrow1.$

In the following, I propose  an amended MaxEnt formulation for  systems
with a displaced  equilibrium, find that  the relevant entropy  in
this setting is the Rényi entropy, interpret the mean constraints,
derive the correct form of solutions, propose numerical procedures
for  estimating   the  parameters   of  the   Tsallis  factor  and
characterize the associated entropies. I will also indicate a
duality between the solutions associated with 
classical and generalized mean constraint.
 Finally  I will  discuss the underlying Legendre structure of generalized thermodynamics associated to this setting.

\section{ The amended MaxEnt formulation}

A key for the apparition of Levy distributions and a
probabilistic justification might be that it seems to appear in the case of
modified, perturbated, or displaced classical Boltzmann-Gibbs equilibrium.
This means that the original MaxEnt formulation \textquotedblleft find the
closest distribution to a reference under a mean constraint\textquotedblright%
\ may be amended by introducing for instance a new constraint that displaces
the equilibrium. The partial or displaced equilibrium may be imagined as an
equilibrium characterized by two references, say $P_{1}$ and $Q$. Instead of
selecting the nearest distribution to a reference under a mean constraint, we
may look for a distribution $P^{\ast}$ simultaneously close to two distinct
references: such a distribution will be localized somewhere \ `between' the
two references $P_{1}$ and $Q$. For instance, we may consider a global system
composed of two subsystems characterized by two prior reference distributions.
The global equilibrium is attained for some intermediate distribution, and the
observable may be, depending on the viewpoint or on the experiment, either the
mean under the distribution of the global system or under the distribution of
one subsystem. This can model a fragmentation process: a system $\Sigma(A,B)$
fragments into $A,$ with distribution $P_{1},$ and $B$ with distribution $Q,$
and the whole system is viewed with distribution $P^{\ast}$ that is some
intermediate between $P_{1}$ and $Q.$ This can also model a phase transition:
a system leaves a state $Q$ \ toward $P_{1}$ and presents an intermediate
distribution $P^{\ast}.$

This can be
stated as: \ find $P^{\ast}$ such that the Kullback-Leibler divergence to $Q,$
$\ D(P||Q)$ is minimum (or equivalently the Shannon $Q$-entropy is maximum),
but under the constraint that $D(P||Q)=D(P||P_{1})+\theta,$ where $\theta$ can
be expressed as a log-likelihood.\ The problem simply writes%
\begin{equation}
\left\{
\begin{array}
[c]{c}%
\min_{P}~D(P||Q)=\min_{P}~\int P(x)\log\frac{P(x)}{Q(x)}dx\\
s.t~\text{\ }\theta=D(P||Q)-D(P||P_{1})=\int P(x)\log\frac{P_{1}(x)}{Q(x)}dx
\end{array}
\right.
\end{equation}
and its solution was given by Kullback \cite[page 39]{kullback-book} as an
illustration of his general theorem on constrained minimization of
$\ D(P||Q)$:%
\begin{equation}
P^{\ast}(x)=\frac{P_{1}(x)^{\alpha}Q(x)^{1-\alpha}}{\int P_{1}(x)^{\alpha
}Q(x)^{1-\alpha}dx}, \label{P-star}%
\end{equation}
which is nothing else but the escort distribution (\ref{escort1}) of nonextensive statistics
\cite{beck-book} (although it is generalized here with reference $Q$). The parameter $\alpha$ is simply the Lagrange parameter
associated to the constraint, and it can be shown that necessarily $\alpha\leq 1$. Clearly, distribution $P^{\ast}$ which is the
geometric mean between $P_{1}$and $Q$ realizes a trade-off, governed by
$\alpha,$ between the two references. By dual attainment, we have
\begin{equation}
\left\{
\begin{array}
[c]{c}%
\min_{P}~D(P||Q)\\
s.t~\text{\ }\theta=D(P||Q)-D(P||P_{1})
\end{array}
\right.  =\sup_{\alpha}\left(  \alpha\theta-\log\left(  \int P_{1}(x)^{\alpha
}Q(x)^{1-\alpha}dx\right)  \right)  . \label{P-star-dual}%
\end{equation}
In this last relation, the term $\log\left(  \int P_{1}(x)^{\alpha
}Q(x)^{1-\alpha}dx\right)  $ is directly proportional to the Rényi divergence (\ref{Renyi}). 

\subsection{Observable mean values}

Observable values are as usual the statistical mean under some distributions.
Depending on the viewpoint, the observable may be a mean under distribution
$P_{1},$ the distribution of an isolated subsystem, or under $P^{\ast},$ the
equilibrium distribution between $P$ and $Q.$ Hence, the problem will be
completed by an additionnal constraint, and a possible approach would be to
select distribution $P_{1}$ by further minimizing the Kullback-Leibler
information divergence $D(P||Q)$, but over $P_{1}(x)$ and subject to the mean
constraint. So, the whole problem writes%
\begin{equation}
K=\left\{
\begin{array}
[c]{c}%
\min_{P_{1}}\left\{
\begin{array}
[c]{c}%
\min_{P}~D(P||Q)=\min_{P}~\int P(x)\log\frac{P(x)}{Q(x)}dx\\
\text{subject to:}~\text{\ }\theta=\int P(x)\log\frac{P_{1}(x)}{Q(x)}dx
\end{array}
\right. \\
\text{subject to: }m=E_{P_{1}}[X]\text{ or }m=E_{P^{\ast}}[X]
\end{array}
,\right.  \label{whole-min-pb}%
\end{equation}
where $E_{P}[X]$ represents the statistical mean under distribution
$P:E_{P}[X]=\int xP(x)dx$. This may be tackled in two steps: first minimize
with respect to $P$ taking into account the mean log-likelihood constraint,
and obtain (\ref{P-star}), and second, minimize with respect to $P_{1}.$
Taking into account (\ref{P-star-dual}), problem (\ref{whole-min-pb}) becomes
\begin{equation}
K=\sup_{\alpha}\left[  \alpha\theta-\left\{
\begin{array}
[c]{c}%
\max_{P_{1}}(\alpha-1)D_{\alpha}(P_{1}||Q)\\
\text{subject to: }m=E_{P_{1}}[X]\text{ or }m=E_{P^{\ast}}[X]
\end{array}
\right.  \right]  \label{whole-min-pb2}%
\end{equation}
and \textit{amounts to the extremization of Rényi information divergence under
a mean constraint}. Therefore, we find that the amended MaxEnt formulation leads to the maximization of Rényi (or equivalently Tsallis) entropy subject to a statistical mean constraint. We can note that the second constraint, $m=E_{P^{\ast}}[X]$ is nothing else but the `generalized expectation' of nonextensive statistics that has here a clear interpretation.

{It is important to note that the minimization of
Kullback-Leibler divergence with respect to $P$ and $P_{1},$ subject to the
two constraints, may not always reduce to the two-steps procedure above.}

% % \subsection{Entropy functionals $\mathcal{F}_{\alpha}^{(1)}(m)$ and
% $\mathcal{F}_{\alpha}^{(\alpha)}(m)$}
% 
% Let us denote by $\mathcal{F}_{\alpha}^{(1)}(m)$ and $\mathcal{F}_{\alpha
% }^{(\alpha)}(m)$ two functions of an observable $m$ defined as the minimum of
% the Rényi information divergence under a mean constraint $m$ :%
% %
% \begin{equation}
% \mathcal{F}_{\alpha}^{(1)}(m)\text{ resp. }\mathcal{F}_{\alpha}^{(\alpha
% )}(m)=\left\{
% \begin{array}
% [c]{c}%
% \min_{P_{1}}D_{\alpha}(P_{1}||Q)\\
% \text{subject to: }m=E_{P_{1}}[X]\text{ resp. }m=E_{P^{\ast}}[X]
% \end{array}
% \right.  , \label{F_def}%
% \end{equation}
% and (\ref{whole-min-pb2}) can be written :%
% \begin{equation}
% K=\sup_{\alpha}\left[  \alpha\theta+(1-\alpha)\mathcal{F}_{\alpha}^{(1\text{
% or }\alpha)}(m)\right]  .
% \end{equation}
% Defined as a \ \ `contraction' of Rényi information divergence (\ref{F_def})
% or of Kullback-Leibler information divergence for a given mean $m$ and a
% log-likelihood constraint $\theta$, functionals $\mathcal{F}_{\alpha}%
% ^{(1)}(m)$ and $\mathcal{F}_{\alpha}^{(\alpha)}(m)$ can be considered as
% level-one entropy functionals. It is possible to prove (proof omitted here)
% that these entropy functionals are nonnegative,
% with an unique minimum at $m_{Q\text{ }}$, the mean of $Q$.
% Furthermore, $\mathcal{F}_{\alpha}^{(1)}(m)$ is strictly convex for $\alpha
% \in\lbrack0,1].$
% 

\section{Solutions to the maximization of Rényi $Q$-entropy}

We now consider the maximization of Rényi $Q$-entropy subject to the
classical mean constraint (C) $m=E_{P_{1}}[X]$ and the generalized mean
constraint (G) $m=E_{P^{\ast}}[X]$ as we obtained in (\ref{whole-min-pb2}).
We first begin by some results on a general `Tsallis' distribution, that 
simplify the derivation of exact solutions (proofs are omitted to save space).

%%%%%%%%%%%%%%%%%%%%%%%%%%%%%%%%%%%%%%% RESULTS %%%%%%%%%%%%%%%%%%%%%%%%%%%%%%%%%%%%
\subsection{Preliminary results}

\begin{definition}
Distribution $P_{\nu}^{\#}(x)$ is defined by:
% below will be of particular
% interest in the remainder of the work:
\begin{equation}
P_{\nu}^{\#}(x)=\left[  \gamma(x-\overline{x})+1\right]  ^{\nu}%
Q(x)e^{D_{\alpha}(P_{\nu}^{\#}||Q)}, \label{eq:defLevy_}%
\end{equation}
on domain $\mathcal{D=D}_{Q}\cap\mathcal{D}_{\gamma},$ where $\mathcal{D}%
_{Q}=\left\{  x:Q(x)\geq0\right\}  $ and $\mathcal{D}_{\gamma}=\left\{
x:\gamma(x-\overline{x})+1\geq0\right\}  .$ In this expression, $\overline{x}$
is either (a) a fixed parameter, say $m$, and $P_{\nu}^{\#}(x)$ is a two
parameters distribution, (b) or some statistical mean with respect to $P_{\nu
}^{\#}(x),$ e.g. its \textquotedblleft classical\textquotedblright\ or
\textquotedblleft generalized\textquotedblright\ mean, and as such a function
of $\gamma.$ Observe that distribution $P_{\nu}^{\#}(x)$ is not necessarily
normalized to one. Associated with $P_{\nu}^{\#}(x)$, we also define a partition function 
\begin{equation}
Z_{\nu}(\gamma,\overline{x})=\int_{\mathcal{D}}\left[  \gamma(x-\overline
{x})+1\right]  ^{\nu}Q(x)dx. \label{eq:PartitionFunction}%
\end{equation}

\end{definition}

% \begin{notation}
% A particular exponent $\nu$ will be considered: $\nu=\xi=\frac{1}{\alpha-1}.$
% For this special value, the following simple equalities will be useful:
% (a) $(\xi+1)=\frac{\alpha}{\alpha-1},$ and (b) $\xi\alpha=(\xi+1)=\frac{\alpha}{\alpha-1}.$
% \end{notation}

\begin{notation}
We will denote by $E_{\nu}\left[  X\right]  $ the statistical mean with
respect to the probability distribution associated with $P_{\nu}^{\#}(x),$ and by $E_{\nu}^{(\alpha)}\left[  X\right] $  the generalized $\alpha-$mean. One can observe that 
in the case of the Levy distribution (\ref{eq:defLevy_}), we have $E_{\nu}^{(\alpha)}\left[  X\right]  =E_{\alpha\nu
}\left[  X\right].$ In the special case $\nu=\pm\xi,$ we obtain $E_{\pm\xi
}^{(\alpha)}\left[  X\right]  =E_{\pm(\xi+1)}\left[  X\right] ,$ because
$\xi\alpha=(\xi+1) = \frac{\alpha}{\alpha-1}.$
\end{notation}

\begin{theorem}
\label{l7} 

The Levy distribution $P_{\xi}^{\#}(x)$ with exponent $\nu=\xi, $
is normalized to one if and only if $\overline{x}=E_{\xi}\left[  x\right]  , $
the statistical mean of the distribution, and 
$D_{\alpha}(P_{\xi}^{\#}||Q)=-\log Z_{\xi+1}(\gamma,\overline{x})=-\log Z_{\xi}(\gamma,\overline{x}).$ \\

In the same way, the Levy distribution
$P_{-\xi}^{\#}(x)$ with exponent $\nu=-\xi,$ is normalized to one if and only
if $\overline{x}=E_{-\xi-1}\left[  x\right]  =E_{-\xi}^{(\alpha)}\left[
x\right],$  the generalized $\alpha-$expectation of the distribution, and
$D_{\alpha}(P_{-\xi}^{\#}||Q)=-\log Z_{-(\xi+1)}(\gamma,\overline{x})=-\log Z_{-\xi}(\gamma,\overline{x}),$ with $\alpha\xi=(\xi+1)$.\\

When $\overline{x}$ is a fixed parameter $m,$ this will be only true for a special
value $\gamma^{\ast}$ of $\gamma$ such that $E_{\xi}\left[  x\right]  =m$ or 
$E_{-\xi}^{(\alpha)}\left[  x\right]  =m$, respectively in the first and second case.
\end{theorem}

\begin{remark}
\label{rem:mapping} Here takes place an important remark on \emph{the mapping
}$\overline{x}$\emph{ }$\leftrightarrow$\emph{ }$\gamma$. Consider the
normalized distribution $P_{_{\xi}}^{\#}(x)$
%\begin{equation}
%P_{_{\xi}}^{\#}(x)=\frac{1}{Z_{\xi}(\gamma,\overline{x})}\left[
%\gamma(x-\overline{x})+1\right]  ^{_{\xi}}Q(x) \label{eq:Pxi}%
%\end{equation}
with $\overline{x}=E_{\xi}\left[  x\right].$ This distribution depends on the
sole parameter $\gamma,$ and $\overline{x}$ is a function of $\gamma.$ But
contrary to the intuition,\emph{ the mapping }$\overline{x}$\emph{
}$\leftrightarrow$\emph{ }$\gamma$\emph{ is not necessarily one to one}.\ This
means that a specified value of the mean $\overline{x}=m$ may correspond to
several values of $\gamma,$ and conversely a specified value of $\gamma$ may
give several different means $\overline{x}.$ \ This can be illustrated through
numerical examples.

\end{remark}

\begin{lemma}
\label{l3} Partition functions $Z_{\xi+1}(\gamma,m)$ and $Z_{-\xi
}(\gamma,m)$ are convex functions of $\gamma.$
\end{lemma}

%%%%%%%%%%%%
\subsection{Solutions}

The solutions to the maximization of Rényi $Q$-entropy subject to the
classical mean constraint (C) $m=E_{P_{1}}[X]$ and the generalized mean
constraint (G) $m=E_{P^{\ast}}[X]$ are found using standard Lagrangian techniques
The optimum solution, see for instance \cite{Boyd2004}, is a
saddle point of the Lagrangian and we may proceed in two steps: first minimize the Lagrangian in $P(x)$, and
thus obtain a solution in terms of the Lagrange parameters, and then maximize
the resulting Lagrangian, the dual function, in order to exhibit the optimum
Lagrange parameters. Taking 
into account the normalization conditions described above, 
these solutions are easily derived and simplified:

\begin{align}
(C)\ \text{\ }\ {P}_{C}{(x)}  &  =\frac{{\left[  \gamma(x-\overline
{x})+1\right]  }^{\xi}}{Z_{\xi}(\gamma,\overline{x})}{Q(x)},\text{ with
}\overline{x}=E_{P_{C}}[X]=E_{\xi}[X]\label{eq:pc}\\
(G)\ \text{\ }P_{G}(x)  &  =\frac{\left(  1+\gamma(x-\overline{x})\right)
^{-\xi}}{Z_{-\xi}(\gamma,\overline{x})}Q(x)\text{ with }\overline{x}%
=E_{P_{G}}[X]=E_{-(\xi+1)}[X] \label{eq:pg}%
\end{align}
where $\xi=\frac{1}{\alpha-1}$ , and $Z_{\nu}(\gamma,\overline{x}%
)$ \ is the partition function. 
It is important to emphasize that $\overline{x}$ in (\ref{eq:pc}) is the
statistical mean with respect to ${P}_{C}{(x),}$ $\overline{x}$ in
(\ref{eq:pg}) is the generalized $\alpha$-mean with respect to ${P}_{G}{(x),}$
and as such a function of $\gamma.$ It is a common mistake in the large majority of reported 
results and calculations to improperly take for $\overline{x}$ the fixed value 
$m$ of the constaint, which is only correct for the optimum value of the Lagrange parameter.

These optimum distributions appear
to be self-referential, since their expressions involve their statistical
mean. Therefore, the direct determination of their parameters is difficult, if
not intractable.

\subsection{Alternate dual functions}

From the Lagrangian theory, one should  maximize
the dual function in order to obtain the remaining Lagrange parameter. 
But in the present cases, the dual functions are implicitely defined. 
Thus, in order to identify the value of the natural parameter associated to the
mean constraints, I propose two \ `alternate' (but effectively computable)
dual functions, whose numerical maximizations enable to exhibit the optimum parameters.

For the classical mean, I just sketch the procedure.  At the optimum, we
have $D(\gamma^{\ast})=\sup_{\gamma}\sup_{\mu}\inf_{P}L(P,\gamma,\mu)$. 
For any value $\widetilde{\mu}$ of $\mu$, letting $\widetilde{D}(\gamma
)=L(P_{\gamma,\widetilde{\mu}}^{\ast},\gamma,\widetilde{\mu}),$ we have 
$D(\gamma^{\ast})\geq\widetilde{D}(\gamma).$ Thus, if $\widetilde{D}%
(\gamma^{\ast})=D(\gamma^{\ast})$ for the optimum $\gamma^{\ast},$ then
$\widetilde{D}(\gamma^{\ast})$ will be a maximum of $\widetilde{D}(\gamma)$
and the maximization of the dual function can be carried equivalently via the
maximization of $\widetilde{D}(\gamma).$  Condition
$\widetilde{D}(\gamma^{\ast})=D(\gamma^{\ast})$ is achieved with $\widetilde{\mu}(\gamma)=-\left(  \xi+1\right)
\left(  1-\gamma m\right).$ Then, after some algebra, 
we obtain the very simple form%
\begin{equation}
\widetilde{D}_{C}(\gamma)=-\log Z_{\xi+1}(\gamma,m)
\end{equation}
that is simply the expression of the divergence from $P_{\xi}^{\#}$ to $Q,$
$D_{\alpha}(P_{\xi}^{\#}||Q)$.
% with
%\begin{equation}
%P_{\xi}^{\#}{(x)}=\frac{{\left[  \gamma(x-m)+1\right]  }^{\xi}}{{Z_{\xi
%}(\gamma,m)}}{Q(x).}%
%\end{equation}
We know that $Z_{\xi+1}(\gamma,m)$ is a convex function. Thus, if $Z_{\xi
+1}(\gamma,m)$ is defined on a continuous domain, $\widetilde{D}_{C}(\gamma)$
has an only maximum for $\gamma=\gamma^{\ast}.$ If $Z_{\xi+1}(\gamma,m)$ is
defined (and convex) on several intervals, $\widetilde{D}_{C}(\gamma)$ may
have a maximum on each of these intervals, and one has to select the minimum
of these maxima (that is the maximum associated with the minimum divergence).
Hence, the identification of the optimum parameter $\gamma^{\ast}$ simply
amounts to the unconstrained maximization of an unimodal functional, possibly
in several intervals.

For the generalized mean, the rationale for an alternate dual function is as follows. 
We know that $D_{\alpha}(P_{-\xi}%
^{\#}||Q)=-\log Z_{-\xi}(\gamma,m)$ when the generalized mean constraint is
satisfied. Since $\frac{d\log Z_{-\xi}(\gamma,m)}{d\gamma}=-\xi\left(
\overline{x}-m\right)  \frac{Z_{-\xi-1}(\gamma,m)}{Z_{-\xi}(\gamma,m)},$ $-\log Z_{-\xi}(\gamma,m)$ is maximum when 
 the constraint $\overline{x}=m$ is satisfied. 
Hence,  the
search of the optimum Lagrange parameter can be carried using the very simple
alternate dual function
\begin{equation}
\widetilde{D}_{G}(\gamma)=-\log Z_{-\xi}(\gamma,m).
\end{equation}
The partition function $Z_{-\xi}(\gamma,m)$ is a convex function for
$\alpha\leq1$. If it is defined on a continuous domain, $\widetilde{D}%
_{G}(\gamma)$ has an only maximum that is simply reached for $\gamma^{\ast}$
such that $m=E_{-\xi-1}[x],$ the generalized $\alpha$-mean$.$ If the domain is
given by several intervals, then $\widetilde{D}_{G}(\gamma)$ may present
several maxima, and the minimum of these maxima, associated with the minimum
divergence $D_{\alpha}(P_{-\xi}^{\#}||Q)$, has to be selected. We thus obtain 
two practical numerical schemes for the identification of the distributions parameters,
and it is also possible to study the behaviour of entropies associated with some 
particular references $Q$.  We come to a close to this presentation by considering the relationship 
between the two minimization problems and an underlying Legendre structure.

%%%%%%%%%%%%%%%%%%%%%%%%%%%%%%%%
\section{Duality and Legendre structure}
\subsection{The $\alpha\leftrightarrow1/\alpha$ duality}

\label{seq:alpha_1_over_alpha_duality}

The dual functions associated to the two problems are $-\log Z_{\xi_{1}%
+1}(\gamma,m)$ and $-\log Z_{-\xi_{2}}(\gamma,m).$ Thus, we will have
pointwise equality of dual functions, and of course of the optima, if $\xi_{1}+1=-\xi_{2},$ that is if
indexes $\alpha_{1}$ and $\alpha_{2}$ satisfy $\alpha_{1}=1/\alpha_{2}.$ 
% In this case, 
% we will have equality of the optimum parameters $\gamma,$ and 
%the two optimization problems will have the same optimum value. 
We can also remark
that with $-\xi_{2}=\xi_{1}+1=\alpha_{1}\xi_{1},$
we have the following relations between the two optimum probability density
functions:
\begin{equation}
P_{G}=\frac{P_{C}^{\alpha_{1}}Q^{1-\alpha_{1}}}{Z_{\xi_{1}}^{1-\alpha_{1}}%
}\text{ \ and \ }P_{C}=\frac{P_{G}^{\alpha_{2}}Q^{1-\alpha_{2}}}{Z_{\xi_{1}%
}^{1-\alpha_{2}}},\text{ \ with \ }\alpha_{2}=1/\alpha_{1},
\end{equation}
and using the fact that $Z_{\xi_{1}+1}(\gamma,m)=Z_{\xi_{1}}(\gamma,m)$ for
the optimum value of $\gamma.$ It means that $P_{G}$ is the escort
distribution of $P_{C}$ with index $\alpha_{1}$ and that $P_{C}$ is the escort
distribution associated with $P_{G}$ and index $\alpha_{2}$. 
% Such interesting
% connection was already noted in \cite[pages 543-544]{Tsallis1998}. Because of
% the pointwise equality of the dual functions, it is clear that the associated
% divergence will be equal at the optimum, that is $D_{\alpha_{1}}%
% (P_{C}||Q)=D_{\alpha_{2}}(P_{G}||Q).$ 
It can be checked in the general case
% Let us consider the $1/\alpha$ Rényi divergence to $Q$ for the escort
% distribution $P^{\ast}(x)=\frac{P_{1}(x)^{\alpha}Q(x)^{1-\alpha}}{\int
% P_{1}(x)^{\alpha}Q(x)^{1-\alpha}dx}$. It writes%
% \begin{align*}
% D_{\frac{1}{\alpha}}(P^{\ast}||Q)  &  =\frac{1}{\frac{1}{\alpha}-1}\log
% \int\left(  P^{\ast}(x)\right)  ^{\frac{1}{\alpha}}Q(x)^{1-\frac{1}{\alpha}%
% }dx\\
% &  =\frac{\alpha}{1-\alpha}\left(  \log\int P_{1}(x)dx-\frac{1}{\alpha}%
% \log\int P_{1}(x)^{\alpha}Q(x)^{1-\alpha}dx\right)  ,
% \end{align*}
% which simplifies immediately to
% \begin{equation}
% D_{\frac{1}{\alpha}}(P^{\ast}||Q)=\frac{1}{\alpha-1}\log\int P_{1}(x)^{\alpha
% }Q(x)^{1-\alpha}dx=D_{\alpha}(P_{1}||Q).
% \end{equation}
% Therefore, we
that  \emph{always} have the equality $D_{\frac{1}{\alpha}}(P^{\ast
}||Q)=D_{\alpha}(P_{1}||Q)$ between the $1/\alpha$ Rényi divergence of the
escort distribution to $Q$ and the standard $\alpha$ divergence 
%(of course, this is also true for Tsallis divergence). 
Hence, the minimization of the
$\alpha$ Rényi divergence subject to the generalized mean constraint is
exactly equivalent to the minimization of the $1/\alpha$ Rényi divergence
subject to the classical mean constraint 
% and the two problems have the same optimum value, %
%
% \begin{equation}
% \left\{
% \begin{array}
% [c]{c}%
% \inf_{P_{1}}D_{\alpha}(P_{1}||Q)\\
% s.t\text{ \ }E_{P^{\ast}}\left[  X\right]  =m
% \end{array}
% \right.  =\left\{
% \begin{array}
% [c]{c}%
% \inf_{P^{\ast}}D_{\frac{1}{\alpha}}(P^{\ast}||Q)\\
% s.t\text{ \ }E_{P^{\ast}}\left[  X\right]  =m
% \end{array}
% \right.  , \label{eq:equality_pbs_1_over_alpha}%
% \end{equation}
so that generalized and classical mean constraints can always be swapped,
provided the index $\alpha$ is changed into $1/\alpha,$ as was argued in
\cite{Raggio1999a,Naudts2002}. 

\subsection{The Legendre structure}
\label{seq:Legendre_structure} 

In the study of alternative entropies, considerable efforts have been directed
to the analysis of associated thermodynamics. 
% Accordingly, this important
% point can not be avoided here. We considered the minimization of $D_{\alpha
% }(P||Q)$ (the maximization of the $\alpha\ Q-$entropy) subject to two
% constraints and ended up with probability density functions (\ref{Csolution})
% and (\ref{Gsolution}). 
The concave entropies  corresponding to our two problems are
$\ S_{C}=\log Z_{\xi+1}(-\frac{\lambda}{\left(  \xi+1\right)},\overline{x})$, 
and $S_{G}=\log Z_{-\xi}(\lambda/\xi,\overline{x})$.
Let us consider the general form $S=\log Z_{\nu+1}(\gamma,\overline{x}).$ 

% By Lemma 13 of Part I,
%\ref{l5_0}
% we have
% \begin{equation}
% \frac{dS}{d\gamma}=\frac{d}{d\gamma}\log Z_{\nu+1}(\gamma,\overline
% {x})=-\gamma\left(  \nu+1\right)  \frac{d\overline{x}}{d\gamma}.
% \end{equation}
In terms of the Lagrange multiplier $\lambda,$ it can be shown that 
\begin{equation}
\frac{dS}{d\lambda}=\frac{dS}{d\gamma}\frac{d\gamma}{d\lambda}=-\gamma\left(
\nu+1\right)  \frac{d\overline{x}}{d\lambda}.
\end{equation}
Specializing the result to the two entropies, we obtain in both
cases the Euler formula:
\begin{equation}
\frac{dS\text{ }}{d\lambda}=\lambda\frac{d\overline{x}}{d\lambda}.
\end{equation}
Next, the derivative of the entropy with respect to the mean is simply
% the Lagrange parameter:
\begin{equation}
\frac{dS\text{ }}{d\overline{x}}=\frac{dS\text{ }}{d\lambda}\frac{d\lambda
}{d\overline{x}}=\lambda\frac{d\overline{x}}{d\lambda}\frac{d\lambda
}{d\overline{x}}=\lambda.
\end{equation}

% This last relation enables us to easily conclude on the unimodality of
% functions $\mathcal{F}_{\alpha}^{(1)}(x)$ and $\mathcal{F}_{\alpha}^{(\alpha
% )}(x)$ that was postponed in Lemma 1 of Part 1. Indeed, $d\mathcal{F}_{\alpha
% }^{(.)}(m)/dx=-dS/dx=-\lambda,$ and functions $\mathcal{F}_{\alpha}^{(.)}(x)$
% are only minimum if $\lambda=0.$ The corresponding optimum probability density
% functions, with $\lambda=0,$ are simply $P_{1}=P^{\ast}=Q,$ and $D_{\alpha
% }(Q||Q)=0.$ Therefore, $\mathcal{F}_{\alpha}^{(1)}(x)$ and $\mathcal{F}%
% _{\alpha}^{(\alpha)}(x)$ have an unique minimum for $x=m_{Q},$ the mean of
% $Q$, and $\mathcal{F}_{\alpha}^{(.)}(m_{Q})=0.$ \newline

Let us now introduce the Massieu potential $\phi(\lambda)=S-\lambda
\overline{x}$ (or equivalently the free energy). Derivations with respect to
the Lagrange parameter and to the mean give%
\begin{equation}
\frac{d\phi\text{ }}{d\lambda}=-\overline{x}, \mbox{~~and~~~} \frac{d\phi\text{ }}{d\overline{x}}=-\overline{x}%
\frac{d\lambda}{d\overline{x}}.
\end{equation}
% 
% \begin{equation}
% \frac{d\phi\text{ }}{d\lambda}=\frac{dS\text{ }}{d\lambda}-\overline
% {x}-\lambda\frac{d\overline{x}}{d\lambda}=-\overline{x},
% \end{equation}
% and%
% \begin{equation}
% \frac{d\phi\text{ }}{d\overline{x}}=\frac{dS\text{ }}{d\overline{x}}%
% -\lambda-\overline{x}\frac{d\lambda}{d\overline{x}}=-\overline{x}%
% \frac{d\lambda}{d\overline{x}}.
% \end{equation}
These four relations show that $S$ and $\phi$ are conjugated with variables
$\overline{x}$ and $\lambda:$ $S$ $\left[  \overline{x}\right]
\rightleftharpoons\phi$ $\left[  \lambda\right]  ,$ so that the basic Legendre
structure of thermodynamics is preserved (but care must be taken for
interpretations, for instance a valid definition of temperature requires that
$\lambda$ always remains positive). 

% The conservation of Legendre structure for
% arbitrary forms of the entropy was enlighted in \cite{Plastino1997} and showed
% to be a direct consequence of Jaynes' Maxent principle, in the case of a
% standard mean constraint. We recover here the Legendre structure for both the
% classical and generalized mean constraint. In fact, given the Legendre
% structure for the classical mean constraint, the Legendre structure for the
% generalized mean simply follows by the $\alpha\leftrightarrow1/\alpha$ duality.

%%%%%%  BIBLIOGRAPHY %%%%%%%%%%%

{\small 
\bibliographystyle{ieeetr}
\bibliography{ree7}

\begin{thebibliography}{10}

\bibitem{Tsallis1999}
C.~Tsallis, ``Nonextensive statistics: Theoretical, experimental and
  computational evidences and connections,'' {\em Brazilian Journal of
  Physics}, vol.~29, pp.~1--35, March 1999.

\bibitem{Tsallis2002}
C.~Tsallis, ``Entropic nonextensivity: a possible measure of complexity,'' {\em
  Chaos, Solitons,\& Fractals}, vol.~13, pp.~371--391, 2002.

\bibitem{Tsallis1988}
C.~Tsallis, ``Possible generalization of {B}oltzmann-{G}ibbs statistics,'' {\em
  Journal of Statistical Physics}, vol.~52, no.~1-2, pp.~479--487, 1988.

\bibitem{LaCour2000b}
B.~R. La~Cour and W.~C. Schieve, ``Tsallis maximum entropy principle and the
  law of large numbers,'' {\em Phys. Rev. E}, vol.~62, pp.~7494 -- 7496,
  November 2000.

\bibitem{Grendar2004}
J.~Grendar, M. and M.~Grendar, ``Maximum entropy method with non-linear moment
  constraints: challenges,'' in {\em Bayesian inference and maximum entropy
  methods in science and engineering} (G.~Erickson, ed.), AIP, 2004.

\bibitem{Beck2004a}
C.~Beck, ``Generalized statistical mechanics of cosmic rays,'' {\em Physica A},
  vol.~331, pp.~173--181, january 2004.

\bibitem{Wilk2000}
G.~Wilk and Z.~Wodarczyk, ``Interpretation of the nonextensitivity parameter q
  in some applications of {T}sallis statistics and {L}évy distributions,'' {\em
  Physical Review Letters}, vol.~84, pp.~2770--2773, March 2000.

\bibitem{Jaynes1963}
E.~T. Jaynes, {\em Statistical Physics}, ch.~Information Theory and Statistical
  Mechanics, pp.~181--218.
\newblock Benjamin, New York, 1963.

\bibitem{kullback-book}
S.~Kullback, {\em Information Theory and Statistics}.
\newblock Wiley, New York, 1959.

\bibitem{beck-book}
C.~Beck and F.~Schloegl, {\em Thermodynamics of Chaotic Systems}.
\newblock Cambridge University Press, 1993.

\bibitem{Boyd2004}
S.~Boyd and L.~Vandenberghe, {\em Convex Optimization}.
\newblock Cambridge University Press, 1st~ed., March 2004.
\newblock ISBN: 0521833787.

\bibitem{Raggio1999a}
G.~A. Raggio, ``On equivalence of thermostatistical formalisms.''
  http://arxiv.org/abs/cond-mat/9909161, 1999.

\bibitem{Naudts2002}
J.~Naudts, ``Dual description of nonextensive ensembles,'' {\em Chaos,
  Solitons, and Fractals}, vol.~13, no.~3, pp.~445--450, 2002.

\end{thebibliography}
}

%%%%%%%%%%%%%%%%%%%%%%%%%%%%%%%%%%%%%%%%%%%%%%%%%%%%%%%%%%%%%%%%%%%%%%%%%%%%%%%%%%%%%%%%%%%%%%%%%

\end{document}